\documentclass[aps,preprint]{revtex4}
\usepackage[dvips]{graphicx}
\begin{document}
\draft
\title{\bf Experimental investigation of electric field distributions in a chaotic 3D microwave rough billiard}
\author{Oleg Tymoshchuk, Nazar Savytskyy, Oleh Hul, Szymon Bauch, and Leszek Sirko}

\address{Institute of Physics, Polish Academy of Sciences, Aleja  Lotnik\'{o}w 32/46, 02-668 Warszawa, Poland}
\date{January 28, 2007}

\bigskip

\begin{abstract}

We present the first experimental study of the electric field
distributions $E_N$  of a three-dimensional (3D) microwave chaotic
rough billiard with the translational symmetry. The translational
symmetry means that the cross-section of the billiard is invariant
under translation along $z$ direction. The 3D electric field
distributions were measured up to the level number $N=489$. In
this way the experimental spatial correlation functions
$C_{N,p}({\bf x,s})\propto \langle E_{N,p}({\bf x}+\frac{1}{2}{\bf
s})E_{N,p}^{\ast}({\bf x} - \frac{1}{2}{\bf s})\rangle$ were found
and compared with the theoretical ones. The experimental results
for higher two-dimensional level number $N_{\bot}$ appeared to be
in good agreement with the theoretical predictions.

\end{abstract}


\bigskip
\maketitle

\smallskip

In this paper we present the first experimental investigation of
electric field distributions  of the chaotic 3D microwave rough
billiard with the translational symmetry. Due to experimental
difficulties there are very few experimental studies devoted to 3D
chaotic microwave cavities
\cite{Sirko1995,Alt1997,Dorr1998,Eckhardt1999,Dembowski2002}. In a
pioneering experiment Deus {\it et al.} \cite{Sirko1995} have been
measured eigenfrequencies of the 3D chaotic (irregular) microwave
cavity in order to confirm that their distribution displays
behavior characteristic for classically chaotic quantum systems,
viz., the Wigner distribution. Three-dimensional chaotic cavities
as well as properties of random electromagnetic vector field have
been also scarcely studied theoretically
\cite{Primack2000,Prosen1997,Arnaut2006}.

In general, there is no analogy between quantum billiards and
electromagnetic cavities in three dimensions. However, for 3D
cavities with the translational symmetry the classification of the
modes into transverse electric (TE) and transverse magnetic (TM)
is possible. The TM modes are especially important because they
allow for the simulation of 2D quantum billiards on
cross-sectional planes of  3D cavities. Furthermore, we show in
this paper that the distributions of the electric field of TM
modes of the 3D chaotic rough cavity can be experimentally
measured.

\begin{figure}[!]
\begin{center}
\rotatebox{0} {\includegraphics[width=0.5\textwidth,
height=0.8\textheight, keepaspectratio]{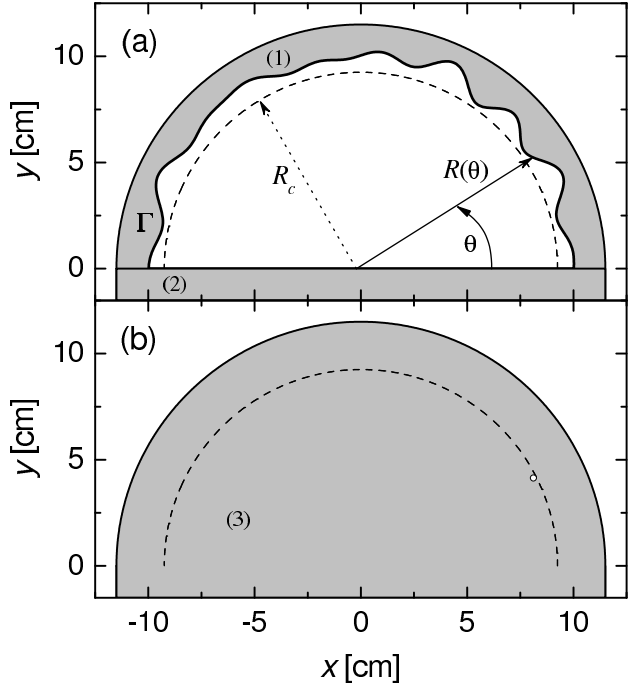}} \caption{Upper
panel: Sketch of  the chaotic half-circular 3D microwave rough
billiard in the $xy$ plane. Dimensions are given in cm. The cavity
sidewalls are marked by 1 and 2 (see text). Squared wave functions
$|\psi_{N,p}(R_c,\theta )|^2$  were evaluated on a half-circle of
fixed radius $R_c=9.25$ cm. Billiard's rough boundary $\Gamma $ is
marked with the bold line. Lower panel: White circle marks the
position of the hole drilled in the upper wall of the cavity. This
hole was used to introduce the perturber inside the cavity in
order to measure the $z$-component of the electric field
distributions $E_{N,p}({\bf x})$.}\label{Fig1}
\end{center}
\end{figure}

In the experiment we used 3D cavity with the translational
symmetry in the shape of a rough half-circle (Fig.~\ref{Fig1})
with the height $h=60$ mm. The cavity was made of polished
aluminium.

We suppose that the direction of the translational symmetry of the
cavity is along the $z$-axis. The boundary conditions at $z=0$ and
$z=h$ demand that the $z$ dependence of the $z$-component of the
electric and magnetic fields $E_{N,p}({\bf x})$ and $B_{N,p}({\bf
x})$  of TM modes be in the form $E_{N,p}({\bf x})\equiv
E_{N,p}(x,y,z)=A_{N,p} \psi_{N,p}(x,y) f_{p}(z)$, where
$f_{p}(z)=\cos(p\pi z/h)$, $p=0,1,2\ldots$, $A_{N,p}$ is the
normalization constant and $B_{N,p}({\bf x})=0$. The functional
dependence of $E_{N,p}({\bf x})$ on the plane cross section
coordinates is denoted by the amplitude $\psi_{N,p}(x,y)\equiv
E_{N,p}(x,y)$. The amplitude $\psi_{N,p}(x,y)$ satisfies the
Helmholtz equation
$$
(\bigtriangleup _{\bot} +k_{N,p}^2)\psi_{N,p}(x,y)=0, \eqno(1)
$$
where $\bigtriangleup _{\bot}$ is two-dimensional Laplacian
operator and $k_{N,p}= (k_N^2-(p\pi/h)^2)^{1/2}$ is the effective
wave vector. The wave vector $k_N=2\pi \nu_N/c$, where $\nu_N$ is
the resonance frequency of the level $N$ and $c$ is the speed of
light in the vacuum. The equation (1) is equivalent to the
Schr\"odinger equation (in units $\hbar =1$) describing a particle
of mass $m=1/2$ with the kinetic energy $k_N^2$ in an external
potential $V=(p\pi/h)^2$ \cite{Kim2005}. Therefore,  microwave 3D
cavities with the translational symmetry simulate on the
cross-sectional planes quantum billiards with the external
potential $(p\pi/h)^2$. In this way microwave cavities can be
effectively used  beyond the standard 2D frequency limit (the case
$p=0$) \cite{Hans} in simulation of quantum systems. The amplitude
$\psi_{N,p}(x,y)$ fulfills Dirichlet boundary conditions on the
sidewalls of the billiard. Therefore, throughout the text the
amplitudes $\psi_N(x,y)$ are also often called the  wave functions
$\psi_N(x,y)$. It is important to note that the full electric
field $E_{N,p}({\bf x})$  satisfies additionally Neumann boundary
conditions at the top and the bottom of the cavity.

Because of the relatively low quality factor of the cavity ($Q
\simeq 4000$) the value of the level number $N$ was evaluated from
the Balian--Bloch formula \cite{Balian}
$$
N(k) = \frac{1}{3\pi ^2}Vk^3 -\frac{2}{3\pi^
2}\int_S\frac{d\sigma_{\omega}}{R_{\omega}}k,\eqno(2)
$$
where k is the wave vector, $V=(9.43 \pm 0.01)\cdot 10^{-4}$ m$^3$
is the volume of the cavity and
$\int_S\frac{d\sigma_{\omega}}{R_{\omega}}=0.932$ m $\pm 0.005$ m
is the surface curvature averaged over the surface of the cavity.

The measurements allowed us for the first time to evaluate the
spatial correlation function \cite{Kaufman88}
$$C_{N,p}({\bf x},{\bf s})= \frac{1}{\langle |E_{N,p}({\bf x})|^2\rangle}
 \langle E_{N,p}({\bf x}+\frac{1}{2}{\bf s})E_{N,p}^{\ast}({\bf x} -
\frac{1}{2}{\bf s})\rangle, \eqno(3)$$ where the local average
$\langle \cdots \rangle$ is defined as follows $$\langle
|E_{N,p}({\bf x})|^2\rangle =
\frac{1}{\Delta^n}\int_{-\Delta/2}^{\Delta/2}| E_{N,p}({\bf
x}+{\bf s})|^2d^ns. \eqno(4)$$

The 3D cavity sidewalls are made of 2 segments (see
Fig.~\ref{Fig1}). The rough segment 1 is described on the
cross-sectional planes by the radius function
$R(\theta)=R_{0}+\sum_{m=2}^{M}{a_{m}\sin(m\theta+\phi_{m})}$,
where the mean radius $R_0$=10.0 cm, $M=20$, $a_{m}$ and
$\phi_{m}$ are uniformly distributed on [0.084,0.091] cm and
[0,2$\pi$], respectively, and $0\leq\theta<{\pi}$. (Here, for the
convenience, the polar coordinates $r$ and $\theta$ are used
instead of the Cartesian ones $x$ and $y$.)

The surface roughness of a billiard on the cross-sectional planes
is characterized by the function $k(\theta)=(dR/d\theta)/R_0$. For
our billiard we have the angle average $\tilde k=(\left
<k^{2}(\theta)\right >_{\theta})^{1/2}\simeq 0.400$. In such a
billiard the classical dynamics is diffusive in orbital momentum
due to collisions with the rough boundary because $\tilde k $ is
much above the chaos border $k_c=M^{-5/2}=0.00056$ \cite{Frahm97}.
The roughness parameter $\tilde k $ determines also other
properties of the billiard \cite{Frahm} on the cross-sectional
planes. The amplitudes $\psi_{N,p}(r,\theta)$ are localized for
the two-dimensional level number $N_{\bot} < N_e = 1/128 \tilde
k^4$, where $N_{\bot} =
\frac{A}{4\pi}k_{N,p}^2-\frac{P}{4\pi}k_{N,p}$. $A=(1.572 \pm
0.002)\cdot 10^{-2}$ m$^{2}$ and $P=0.537$ m  $\pm 0.001$ m are
the cross-sectional plane area and its perimeter, respectively.
Because of a large value of the roughness parameter $\tilde k $
the localization border lies very low, $N_e \simeq 1$. The border
of Breit-Wigner regime is $N_W = M^2/48\tilde k^2 \simeq 52$. It
means that between $N_e < N_{\bot} < N_W$ Wigner ergodicity
\cite{Frahm} ought to be observed and for $N_{\bot} > N_W$
Shnirelman ergodicity should emerge.

To measure the amplitudes $\psi_{N,p}(r,\theta)$ of the 3D
electric field distributions we used a  very effective  method
described in \cite{Savytskyy2003}. It is based on the perturbation
technique \cite{Slater52} and preparation of the ``trial
functions" \cite{Savytskyy2004,Hul2005,Hul2006}. In the
perturbation method a small perturber is introduced inside the
cavity to alter its resonant frequencies and in this way to
evaluate the squared wave functions $|\psi_{N,p}(R_c, \theta )|^2$
(see Fig. ~\ref{Fig1}). The perturber (4.0 mm in length and 0.3 mm
in diameter, oriented in $z$-direction) was moved by the stepper
motor via the Kevlar line hidden in the groove (0.4 mm wide, 1.0
mm deep) made in the cavity's bottom wall along the half-circle
$R_c$. The measurements were performed at 0.36 mm steps along a
half-circle with fixed radius $R_c=9.25$ cm.

In order to find the dependence of the electric field
distributions $E_{N,p}({\bf x})$ on the $z$ coordinate and to
estimate the wave vector $k_3=p\pi/h$ we measured the electric
field inside the 3D cavity along the $z$-axis. Also in this case
the perturber (4.5 mm in length and 0.3 mm in diameter) was
attached to the Kevlar line and moved by the stepper motor. The
perturber entered and exited the cavity by small holes (0.4 mm)
drilled in the upper and the bottom walls of the cavity. The both
holes were located at the position: $r=9.11$ cm, $\theta=0.47$
radians.

Using the method of the ``trial wave function" we were able to
reconstruct 75 experimental wave functions $\psi_{N,p}(r, \theta
)$, which belonged to TM modes of the rough half-circular 3D
billiard with the level number $N$ between 2 and 489. The range of
corresponding eigenfrequencies was from $\nu_{2} \simeq 2.47$ GHz
to $\nu_{489} \simeq 11.99$ GHz. The remaining wave functions
belonging to TM modes, from the range $N=2-489$,  were not
reconstructed because of near-degeneration of the neighboring
eigenfrequencies or due to the problems with the measurements of
$|\psi_{N,p}(R_c, \theta )|^2$ along a half-circle coinciding for
its significant part with one of the nodal lines of $\psi_{N,p}(r,
\theta )$.

\begin{figure}[!]
\begin{center}
\rotatebox{0} {\includegraphics[width=0.5\textwidth,
height=0.8\textheight, keepaspectratio]{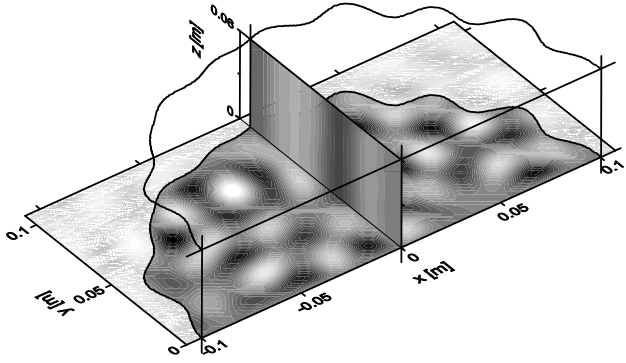}} \caption{ The
reconstructed wave function $\psi_{460,0}(r,\theta) $ of the
chaotic half-circular microwave rough billiard. The amplitudes
have been converted into a grey scale with white corresponding to
large positive and black corresponding to large negative values,
respectively. Dimensions of the billiard are given in cm. In the
figure the $z$ dependence of the electric field distribution $
E_{460,0}(r,\theta,z )|_{x=0} \propto
\psi_{460,0}(r,\theta)|_{x=0} f_{0}(z)$ is also shown.
}\label{Fig2}
\end{center}
\end{figure}

\begin{figure}[!]
\begin{center}
\rotatebox{0} {\includegraphics[width=0.5\textwidth,
height=0.8\textheight, keepaspectratio]{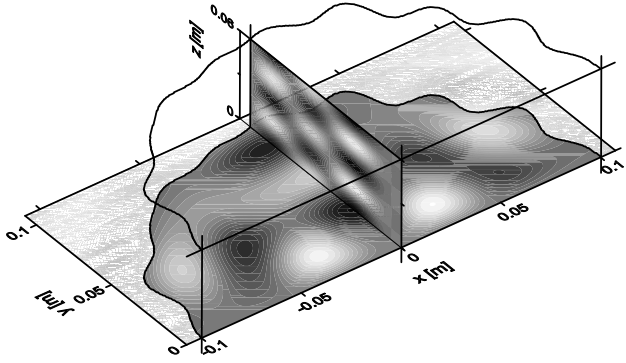}} \caption{ The
reconstructed wave function $\psi_{463,4}(r,\theta) $ of the
chaotic half-circular microwave rough billiard. The $z$ dependence
of the electric field distribution $ E_{463,4}(r,\theta,z )|_{x=0}
\propto \psi_{463,4}(r,\theta)|_{x=0} f_{4}(z)$ is also shown.
}\label{Fig3}
\end{center}
\end{figure}

In Fig. \ref{Fig2} and Fig. \ref{Fig3} we show two examples of
reconstructed wave functions  $\psi_{460,0}(r, \theta )$ and
$\psi_{463,4}(r, \theta )$, respectively. The character of the
wave functions predominantly depends on the effective wave vector
$k_{N,p}$. It is seen that the wave function $\psi_{463,4}(r,
\theta )$ in Fig. \ref{Fig3} is more regular than the one
presented in Fig. \ref{Fig2}  in spite of having the larger level
number $N$.

In order to check ergodicity of the billiard's wave functions
$\psi_{N,p}(R_c, \theta )$, especially close to the ergodicity
borders, one should use some additional measures  such as e.g.,
calculation of the structures of their energy surfaces
\cite{Frahm97}.  For this reason we extracted wave function
amplitudes $C^{(N,p)}_{nl}=\left< n,l|N,p \right>$ in the basis
$n, l$ of a half-circular billiard with radius $r_{max}$, where $
n=1,2,3 \ldots$ enumerates the zeros of the Bessel functions and $
l=1,2,3 \ldots$ is the angular quantum number. As expected, close
to the border of the regimes of Breit-Wigner and Shnirelman
ergodicity the wave function $\psi_{460,0}(r, \theta )$ ($N_{\bot}
= 65$) was found to be extended homogeneously over the whole
energy surface \cite{Hlushchuk01} (figure not shown here). In
contrary, the wave function $\psi_{463,4}(r, \theta )$, $N_{\bot}
= 16$, which lies closer to the localization boarder, displays the
tendency to localization in $n, l$ basis (figure not shown here).

The measurement of 3D electric field distributions $E_{N,p}({\bf
x})$ allowed us for the first time to find the experimental
spatial correlation function $C_{N,p}({\bf x,s})$. It is easy to
show \cite{Berry77} that for the 3D chaotic cavity with the
translational symmetry the spatial correlation function should
have the following form:
$$
C_{N,p}({\bf x},|{\bf s}|)\equiv C_{N,p}(|{\bf
s}|)=J_0(k_{N,p}s_{xy})\cos(p\pi s_z/h), \eqno (5)
$$
where $|{\bf s}|=(s_{xy}^2+s_z^2)^{1/2}$. For the cross-sectional
planes $z=const$ the correlation function $C_{N,p}(|{\bf s}|)\sim
J_0(k_{N,p}s_{xy})$ is reduced to the well known result of Berry
\cite{Berry77} for chaotic 2D wave functions described by a random
superposition of plane waves.

\begin{figure}[!]
\begin{center}
\rotatebox{0} {\includegraphics[width=0.5\textwidth,
height=0.8\textheight, keepaspectratio]{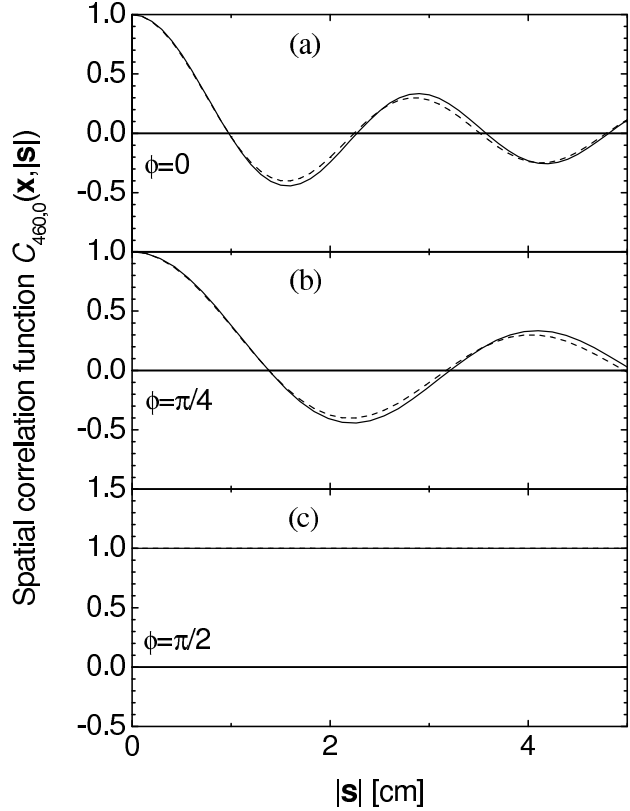}} \caption{Panels
(a)-(c) show the experimental correlation function $C_{460,0}({\bf
x},|{\bf s}|)$ calculated at ${\bf x}=$ (-2.75 cm, 4.35 cm, 0 cm)
for the three projection angles $\phi=0$, $\pi/4$, and $\pi/2$,
respectively. Experimental correlation function $C_{460,0}({\bf
x},|{\bf s}|)$ (full line) is compared with the theoretical one
(dashed line). \label{Fig4}}
\end{center}
\end{figure}

In Fig. \ref{Fig4}(a)-(c) we show a representative example of the
experimental correlation function $C_{460,0}({\bf x},|{\bf s}|)$
($N_{\bot} = 65$) calculated at ${\bf x}=$ (-2.75 cm, 4.35 cm, 0
cm) for the three different projection angles $\phi=0$, $\pi/4$,
and $\pi/2$, respectively, where $\phi = \arcsin(s_z/|{\bf s}|)$.
The local average $\langle \cdots \rangle$ required for the
evaluation of $C_{N,p}({\bf x},|{\bf s}|)$ (see the formulas (3)
and (4)) was calculated on the cross-sectional plane $xy$ in the
range $\Delta/2=2\pi/k_{N,p}$. The experimental correlation
functions $C_{460,0}({\bf x},|{\bf s}|)$ are compared in Fig.
\ref{Fig4} with the theoretical ones. In all cases we find good
agreement with the theoretical predictions given by the formula
(5). Small discrepancies observed in Fig. \ref{Fig4}(a)  for
$|{\bf s}|>1$ can be connected with the finiteness of the system
and were theoretically studied in \cite{Baecker02}.

\begin{figure}[!]
\begin{center}
\rotatebox{0} {\includegraphics[width=0.5\textwidth,
height=0.8\textheight, keepaspectratio]{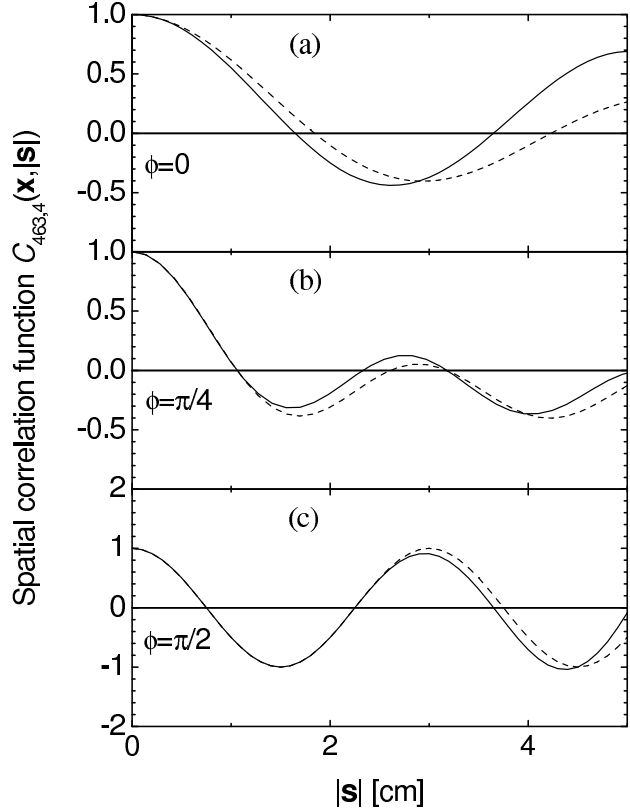}} \caption{Panels
(a)-(c) show the experimental correlation function $C_{463,4}({\bf
x},|{\bf s}|)$ calculated at ${\bf x}=$ (-2.75 cm, 4.35 cm, 0 cm)
for the three projection angles $\phi=0$, $\pi/4$, and $\pi/2$,
respectively. Experimental correlation function $C_{463,4}({\bf
x},|{\bf s}|)$ (full line) is compared with the theoretical one
(dashed line). \label{Fig5}}
\end{center}
\end{figure}

Fig. \ref{Fig5}(a)-(c) shows the experimental correlation function
$C_{463,4}({\bf x},|{\bf s}|)$ ($N_{\bot} = 16$) calculated at
${\bf x}=$ (-2.75 cm, 4.35 cm, 0 cm) for the three different
projection angles $\phi=0$, $\pi/4$, and $\pi/2$, respectively.
The experimental correlation functions $C_{463,4}(|{\bf s}|)$ are
compared in Fig. \ref{Fig5} with the theoretical ones.  In  Fig.
\ref{Fig5} (a), even for small $|{\bf s}|$, we find a significant
departure of the experimental correlation function from the
theoretical prediction, which clearly suggests that the wave
function $\psi_{463,4}(r, \theta )$ is not chaotic. Also in Fig.
\ref{Fig5} (b)-(c)  the experimental correlation functions
$C_{463,4}(|{\bf s}|)$ show for larger $|{\bf s}|$ significant
deviations from the theoretical ones. The discrepancies between
the correlation function $C_{463,4}({\bf x},|{\bf s}|=z)$ for the
$z$-component of the electric field distribution and the
theoretical prediction in Fig. \ref{Fig5}(c) arise mainly due to
the procedure of averaging of the correlation function
$C_{N,p}({\bf x},|{\bf s}|)$ in $z$-direction, which was taken
over the period of the cosine function.

In summary, we measured the wave functions of the chaotic 3D rough
microwave billiard with the translational symmetry up to the level
number $N=489$. For the first time the experimental correlation
function $C_{N,p}({\bf x},{\bf s})$ was estimated and compared
with the theoretical prediction. For the states with higher
$N_{\bot}$ we find, especially for small values of the parameter
$|{\bf s}|$, good agreement with the theoretical predictions,
which show that the wave functions are chaotic. For the states
with lower $N_{\bot}$ significant discrepancies between
experimental and theoretical results are observed.

Acknowledgments.  This work was partially supported by the
Ministry of Science and Higher Education grants No. N202 099
31/0746 and 2 P03B 047 24.

\end{document}